\documentclass[aps, a4paper, showpacs, twocolumn, longbibliography notitlepage, superscriptaddress,10pt]{revtex4-2}
\usepackage[colorlinks=true,linkcolor=blue,urlcolor=blue,citecolor=blue]{hyperref}

\usepackage{textcomp,enumerate,color}                    			
\usepackage[margin=0.8in]{geometry}
\usepackage{tcolorbox}
\usepackage{amsmath,amssymb,amsthm,bbm,bm,nicefrac,}   				
\usepackage{graphicx,epsfig}          
\usepackage[normalem]{ulem}  
\usepackage[sort&compress]{natbib}

\newcommand{\ket}[1]{\left|#1\right\rangle}							
\newcommand{\bra}[1]{\left\langle#1\right|}
\newcommand{\proj}[1]{\ket{#1}\!\!\bra{#1}}
\newcommand{\ketbra}[2]{\left|#1\rangle\langle#2\right|}
\newcommand{\braket}[2]{\left\langle #1\lvert#2\right\rangle}

\newcommand{\mean}[1]{\left\langle #1\right\rangle}

\newcommand{\be}{\begin{equation}} 							
\newcommand{\ee}{\end{equation}}
\newcommand{\ba}{\begin{align}}
\newcommand{\ea}{\end{align}}
\newcommand{\bematrix}{\left(\begin{matrix}}
\newcommand{\ematrix}{\end{matrix}\right)}

\def\one{{\mbox{$1 \hspace{-1.0mm}  {\bf l}$}}}

\def\Z{\ensuremath{\mathbbm{Z}}}
\def\ii{\mathrm{i}}
\def\tr{\mathrm{Tr}}

\def\cB{\mathcal B}

\def\cE{\mathcal E}
\def\cF{\mathcal F}
\def\cG{\mathcal G}
\def\cH{\mathcal H}											
\def\cI{\mathcal I}

\def\cM{\mathcal M}

\def\cS{\mathcal S}
\def\cT{\mathcal T}

\def\cV{\mathcal V}

\newcommand\defn[1]{\textsl{#1}} 								

\begin{document}
\title{Identification of malfunctioning quantum devices}

\author{Michalis Skotiniotis}
\affiliation{Departamento de Electromagnetismo y Física de la Materia, Universidad de Granada, 18010 Granada, Spain}
\affiliation{Institute Carlos I for Theoretical and Computational Physics, Universidad de
Granada, 18010 Granada, Spain}
\affiliation{F\'isica Te\`orica: Informaci\'o i Fen\`omens Qu\`antics, Departament de F\'isica, Universitat Aut\`onoma de Barcelona, 08193 
Bellatera (Barcelona) Spain}
\author{Santiago Llorens}
\affiliation{F\'isica Te\`orica: Informaci\'o i Fen\`omens Qu\`antics, Departament de F\'isica, Universitat Aut\`onoma de Barcelona, 08193 
Bellatera (Barcelona) Spain}    
\author{Ronja Hotz}
\affiliation{Freie Universit\"at Berlin, 14195 Berlin, Germany}

\author{John Calsamiglia}
\affiliation{F\'isica Te\`orica: Informaci\'o i Fen\`omens Qu\`antics, Departament de F\'isica, Universitat Aut\`onoma de Barcelona, 08193 
Bellatera (Barcelona) Spain}
\author{Ramon Mu\~noz-Tapia}
\affiliation{F\'isica Te\`orica: Informaci\'o i Fen\`omens Qu\`antics, Departament de F\'isica, Universitat Aut\`onoma de Barcelona, 08193 
Bellatera (Barcelona) Spain}
\date{\today}
\begin{abstract}
We consider the problem of correctly identifying a malfunctioning quantum device that forms part of a network of $N$ such devices, 
which can be considered as the quantum analogue of classical anomaly detection.  In the case where the devices in question are 
sources assumed to prepare identical quantum pure states, with the faulty source producing a different anomalous pure state, we 
show that the optimal probability of successful identification requires a global quantum measurement. We also put forth several 
local measurement strategies---both adaptive and non-adaptive, that achieve the same optimal probability of success in the limit 
where the number of devices to be checked are large. In the case where the faulty device performs a known unitary operation we 
show that the use of entangled probes provides an improvement that even allows perfect identification for values of the unitary 
parameter that surpass a certain threshold.  Finally, if the faulty device implements a known qubit channel we find that the 
optimal probability for detecting the position of rank-one and rank-two Pauli channels can be achieved by product state inputs and 
separable measurements for any size of network, whereas for rank-three and general amplitude damping channels optimal 
identification requires entanglement with $N$ qubit ancillas.
\end{abstract}
\maketitle
\section{Introduction} 
\label{sec:introduction}
Recent advancements in quantum technologies, such as quantum computing 
devices~\cite{Nigg14, Taminiau14, Corcoles15, Kelly15, Nafradi16, Cramer16}, quantum communication~\cite{Yin17, Ren17, Liao17}, and quantum 
sensors~\cite{Reiter17}, lend credence to the notion that one day soon such devices will be readily available and, hopefully, part of an interconnected quantum network~\cite{Kimble08}.  In turn, the existence of such quantum networks 
gives rise to new technical challenges, such as the correct identification of possible malfunctions. In a vast network of quantum devices---be they sources that produce quantum states, quantum channels that transmit information, or the vast 
array of gates in quantum computers---it is imperative that we are able to find efficient ways to identify faulty components.

\begin{figure}[htb]
\includegraphics[keepaspectratio, width=8cm, height=6cm]{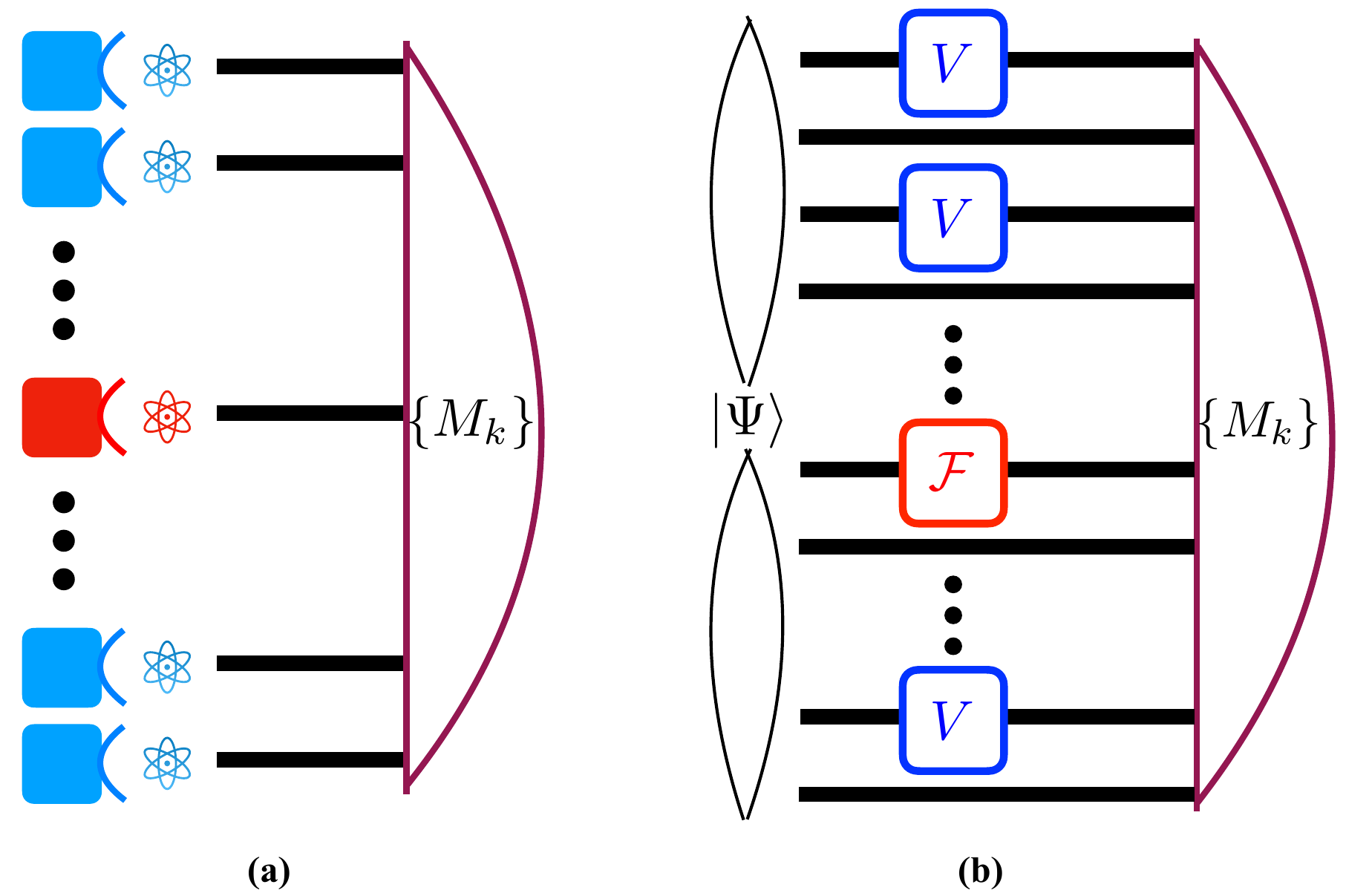}
\caption{Position error identification for sources and channels. (a) $N$ quantum sources are programmed to produce a given state 
$\ket{0}$ except for a faulty device (depicted in red) which produces a known pure state $\ket{\phi}$. (b) $N$ quantum 
gates are programmed to perform a given unitary $V$ except for a faulty device (depicted in red) which may perform a different unitary $W$ or a noisy channel $
\cF$. Whilst in (a) we only have to optimize over all possible 
measurement strategies, in (b) we can also optimize over all possible input state strategies, including those that make use of $N$ 
additional ancillas.}
\label{fig:EPI}
\end{figure}

The identification of rare events that differ significantly from the majority of all other observations is known as anomaly detection and is a fundamental primitive in 
classical data analysis and signal processing~\cite{Tartakovsky2014} with 
a wide range of applications from the identification of  denial of service attacks~\cite{Thottan2003, Siris2004,Tartakovsky2013},
to the identification of fraudulent financial activity~\cite{Ahmed2016} (for a survey of anomaly detection see~\cite{Chandola2009}).
Two important algorithms for anomaly detection, namely kernel principal component analysis and support vector machines,
have been shown to be efficiently applicable in detecting anomalies in quantum data~\cite{Lu:17}.  Here we consider a more 
direct application of anomaly detection where the anomaly is not restricted to classical data but in the most general quantum device.
Specifically, given $N$ \defn{identical} devices programmed to perform a particular task, with one of the devices developing a known 
malfunction, our goal is to optimally identify the malfunctioning device whilst only allowed to query the network once 
(see Fig.~\ref{fig:EPI}). This task that we term \emph{Error Position Identification}, or EPI for short, is a key ingredient in pulse-position 
modulated quantum communication~\cite{Dolinar:06,Dalla:15}, quantum illumination~\cite{Lloyd08}, quantum reading~\cite{Pirandola11}, and target 
detection~\cite{Zhuang17}.

We note that instances of EPI have appeared elsewhere in the literature under the name of channel position finding~\cite{Zhuang20a, Zhuang20b, Harney21, Pereira21} 
where the task is to identify the position of a target bosonic channel, with a given reflectivity, among $m$ background bosonic channels of a different reflectivity. The 
case of identifying the position of a singular unitary channel was also considered recently in~\cite{Hillery23}. Here, we provide a more comprehensive analysis of EPI by 
considering more general devices---including states, unitary gates, and quantum channels---using a variety of different strategies employing probes in separable, 
entangled, as well as ancilla assisted strategies. Unlike~\cite{Zhuang20a, Zhuang20b, Harney21, Pereira21} we consider channels acting on two-dimensional systems, and 
consider both rank-one and rank-two Pauli noise channels, depolarizing, and amplitude damping channels. Moreover, our aim is to identify the position of the 
erroneous device by querying the network only once, as opposed to the more general multi-querying adaptive strategies considered in~\cite{Zhuang20a}.  
For the case of unitary channels we show that there exists instances for which the position of the unitary channel can be identified with certainty, a result that was 
missed in~\cite{Hillery23}. Note that EPI is fundamentally different from the anomaly detection scenario considered in~\cite{Lu:17} which focused in classification of 
states using quantum machine learning techniques.

We first address the setting when the devices are quantum sources known to produce a specific pure state in $\cH_d$ with the faulty source producing a different 
anomalous pure quantum state (the multi-anomaly case has recently been studied in \cite{Llorens23}). We show that the maximum probability of success is achieved by a 
global measurement strategy given by the so-called \defn{square root measurement}, which yields a success probability that converges to a non-zero constant as $N$ 
increases. We next show that when the devices are known to perform a specific unitary gate, with the anomalous device performing a different unitary, a probe state 
with entanglement among the $N$ input states---that can be effectively taken to be qubits---achieves the maximal probability of success without the need of using 
ancillary systems. In fact, we show that for certain non-trivial errors the optimal probability of success can even reach unity. Whilst the use of entangled probe states 
does yield an improvement over strategies employing separable probe states, this improvement diminishes with the size of the network.

We then address the case when the devices are qubit channels and show that
for rank-1 and rank-2 Pauli channels the maximal probability of success is achievable by preparing probes and measuring their 
outputs in a common, suitably chosen local basis. For rank-3 Pauli channels and amplitude damping, we show that one benefits most from 
appropriately entangling the $N$ probes with an additional $N$ ancillas.  

Finally, for the case where the quantum devices are sources we provide several local strategies, both adaptive 
and non-adaptive, whose performance is nearly optimal which is of particular interest for realistic applications.  

\section{Background.}  
\label{sec:background}
The task of successfully identifying the position of a faulty device is equivalent to 
optimally discriminating among $N$ quantum states, in the case of source EPI, or $N$ quantum channels.  For the remainder of this 
work we will be concerned with devices that either prepare systems in a known pure state (sources), or perform a known quantum 
operation (channels). Without loss of generality a faultless quantum source prepares systems in the state $\ket{0}$ whereas the  faulty source---equally likely to be 
located at any of the $N$ positions---prepares the $k^\mathrm{th}$ system in some state 
$\ket{\phi}$. Similarly, a faultless quantum operation performs the unitary operation $V$ whereas the anomalous operation 
performs a completely-positive, trace-preserving (CPTP) map, $\cF^{(k)}$, on the $k^\mathrm{th}$ quantum system~\footnote{In the language of~\cite{Lu:17} EPI is 
an instance of \emph{point} anomaly and, moreover, can be cast in the framework of supervised learning as we assume that the description of both normal and anomalous 
devices is known}.  The probability of successfully identifying the position of the faulty device is given by 
\be
P_S=\frac{1}{N}\sum_{k=1}^Nq(k|k),
\label{eq:prob_succ}
\ee
where $q(l|k)$ is the conditional probability given by $\tr\left[M_l\,\proj{\phi_k}\right]$ in the case of sources, and  
$\tr\left[M_l\,\cE^{(k)}[\proj{\psi}]\right]$ in the case of channels, with  \mbox{$\{M_l\geq 0\, |\sum_{l}M_l=\one\}:=\cM$}
forming a general positive operator-valued measure (POVM). Observe that for source EPI, the optimization over all measurement strategies of Eq.~\ref{eq:prob_succ} is 
a semi-definite program (SDP). For channel EPI, however, Eq.~\ref{eq:prob_succ} needs to be optimized both over the input state as well as the measurement. Luckily 
enough, the formalism of quantum testers~\cite{Chiribella08} allows one to linearize the problem and formulate it as an SDP---as long as the optimization is carried over 
strategies that make use of idler or ancillary systems as shown in Fig.~\ref{fig:EPI}.

By and large analytical solutions to either state or channel discrimination problems are very difficult and known 
only for two~\cite{Holevo73,Helstrom76} or three~\cite{Ha13} pure or mixed states or between states possessing a certain 
symmetry~\cite{Yuen75, Ban:97, Eldar:01, Eldar:04, Barnett:01, Andersson02, Sentis16, Sentis:17,Sentis:18} 
(see also~\cite{Barnett09,Bae:15}). For channel discrimination analytic results are known 
for distinguishing among a finite number of unitaries in a single or finite number of runs~\cite{Acin:01b,Duan:07,Duan:09}, or between 
two arbitrary CPTP maps~\cite{Sacchi:05a, Sacchi:05b, Piani:09}.  We stress that here, unlike most channel discrimination instances 
to date, the identity of the channel (unitary or otherwise) is known, and what one is looking for is the position at which the channel is 
acting.

\section{Results}
\label{sec:results}
\subsection{Source EPI} 
\label{sec:sourceEPI}
We first consider the case of successfully identifying the location of a faulty source as in Fig.~\ref{fig:EPI}(a).
The problem simplifies to optimally discriminating among the set of $N$ linearly independent states 
\be
\ket{\psi_k}=\ket{0}^{\otimes (k-1)}\otimes\ket{\phi}\otimes\ket{0}^{\otimes(N-k)},\quad k\in(1,\ldots,N)\, .
\label{eq:statediscrimination}
\ee
Notice that the set of states also enjoys cyclic symmetry, which allows to  
work out the solution explicitly~\cite{Ban:97, Eldar:01, Eldar:04, Dalla:15,Llorens23}.  

Let $\{\ket{m_j}\}_{j=1}^N$ be an orthonormal basis for the $N$-dimensional 
space spanned by $\{\ket{\psi_k}\}_{k=1}^N$.  Then the probability of success is given by~\cite{Sentis16}
\be
P_S=\frac{1}{N}\sum_{k=1}^N\left|\braket{\psi_k}{m_k}\right|^2=\sum_{k=1}^N\left|B_{kk}\right|^2,
\label{eq:prob_succ_linear_indep}
\ee
where $B$ is but one of an infinitude of square roots of the \defn{Gram} matrix---the matrix of overlaps 
$G_{kl}=\frac{1}{N}\braket{\psi_k}{\psi_l}$, which contains all the relevant information to assess the discrimination properties of a set of quantum states.  As $G>0$, 
the optimization of Eq.~\eqref{eq:prob_succ_linear_indep} is achieved by maximizing over all polar decompositions of $B=VS$, i.e., 
\be
P_S=\max_{V}\sum_{k=1}^N\left|(VS)_{kk}\right|^2 
\label{eq:prob_succ_polar}
\ee
where $S$ is the unique, self-adjoint square root of $G$.  The corresponding measurement is simply given by 
$\{\ket{m_k}=\frac{(VS)^{-1}}{\sqrt{N}}\ket{\psi_k}\}$ (see~\cite{Sentis16}).

For the set in Eq.~\eqref{eq:statediscrimination} the Gram matrix can be easily shown to be
\be
G=\frac{1-b}{N}\one+b \proj{\bm 1},
\label{eq:gram_matrix_source}
\ee
where $b=\braket{\psi_k}{\psi_l}$ for $k\neq l$ and $\ket{\bm 1}=\frac{1}{\sqrt{N}}(1,\ldots,1)^T$. 
The matrix $G$ and has two distinct eigenvalues, $\lambda_1=\frac{1+(N-1)b}{N}$, and $\lambda_2=\frac{1-b}{N}$ where the latter 
is $(N-1)$-fold degenerate. Note that $G$ is circulant~\cite{Horn:12} as is any function of $G$, in particular $S=\sqrt{G}$ whose 
diagonal entries are all equal to
\be
S_{kk}=\frac{\tr S}{N}=\sum_{j=1}^N \frac{\sqrt{\lambda_j}}{N}; \ \ \forall \, k=1,2,\ldots, N.
\label{eq:diag_S}
\ee
The condition that $S_{kk}=S_{ll},\, \forall k\neq l$ is necessary and sufficient to show that Eq.~\eqref{eq:prob_succ_linear_indep} is 
maximized by $B=S$~\cite{Dalla:15, Sentis16}, i.e., 
$P_S^*= N S_{kk}^2=(\tr{S})^2/N$, which reads
\be
P_S^*=\left(\frac{\sqrt{ 1+(N-1) b}
                    +(N-1)\sqrt{1-b}}{N}\right)^2.
    \label{eq:max_prob_succ}
\ee
To compare with the results of next subsection, it is convenient to write the $b$ in terms of the angle $\phi$ defined from the overlap between the default and 
mutated state as $b=|\braket{0}{\phi}|^2=\cos^2\nicefrac{\phi}{2}$.  We get
\be
P_S^*=\left(\frac{\sqrt{1+(N-1)\cos^2\nicefrac{\phi}{2}}+(N-1)\sin\nicefrac{\phi}{2}}{N}\right)^2.
\label{eq:max_prob_succ-2}
\ee

The measurement that achieves this optimal value is the so-called  \defn{square-root} 
measurement~\cite{Ban:97, Eldar:01, Eldar:04, Dalla:15}. Notice that $P_S^*=1$ if and only if $b=0$, i.e., $\ket{\phi}=\ket{1}$. For large strings of states, the success 
probability is finite and reads 
\be
    \begin{split}
P_S^*&=(1-b) +\frac{2\sqrt{b(1-b)}}{\sqrt{N}}+O\left(\frac{1}{N}\right)\\
&=\sin^2\nicefrac{\phi}{2}+\frac{\sin\phi}{\sqrt{N}}+O\left(\frac{1}{N}\right).
    \end{split}
\label{eq:max_prob_asym}
\ee

\subsection{Unitary EPI} 
\label{sec:unitaryEPI}
Let us now consider the case of successfully identifying the location of a faulty unitary gate $W$, i.e.,  
$\cF(\cdot)=W(\cdot) W^\dagger$ in Fig.~\ref{fig:EPI}(b). As the both $V$ and $W$ are known---only the location of the latter
is unknown---we can assume without loss of generality that the states to be discriminated have the form
\be
\begin{split}
\ket{\psi_k}&=\left(\one^{\otimes (k-1)}\otimes U \otimes\one^{\otimes (N-k)}\right)_p\otimes \one_{a}\ket{\psi}\\
&:= U_p^{(k)}\otimes \one_{a}\ket{\psi},\,\,\, k\in(1,\ldots, N),
\end{split}
\label{eq:unitary_discrimination}
\ee
where $U=V^\dagger W$, and $\ket{\psi}\in\cH_p\otimes \cH_a$ is a probe state which in principle can contain ancillary systems. However, we note that the use of 
ancillas here is redundant since as for any probe-plus-ancilla state there exists an $N$ probe state that gives the same Gram matrix~\cite{supple}. Observe that if 
$\ket{\psi}=\ket{\gamma}^{\otimes N}$ with $U\ket{\gamma}=\ket{\phi}$, we recover the source scenario 
discussed above with  success probability given by Eq.~\eqref{eq:max_prob_succ}. The question is whether choosing a more suitable initial state, 
perhaps involving entanglement, improves the probability of success.
  
Using the symmetry of the problem, for any given input state $\rho\in\cB(\cH^{\otimes N})$ 
and optimal POVM $\{M_\sigma\}$ there exists a permutationally invariant state 
$\tilde{\rho}\in\cB(\cH^{\otimes N})$ and permutationally covariant POVM $\{\tilde{M}_\sigma\}$ 
that achieves the same probability of success~\cite{supple}. On the other hand, as the probability of success is 
a convex function the optimal input state can always be chosen to be pure. 
A seemingly natural, though unproven, assumption is to restrict the optimization over states to those that are both pure and permutationally 
invariant. For $N>2$ this assumption amounts to searching over pure states that belong to the 
totally symmetric subspace of $\cH^{\otimes N}$ (see Eq.~\ref{eq:Dicke_states}).  We have numerically 
verified that this is indeed the case for small values of $N$.

As $U$ is known we may write any permutationally invariant probe state with respect to the eigenbasis of $U$, with all amplitudes taken to be real and positive without 
loss of generality. Furthermore, an optimal probe state only involves those eigenstates of $U$ whose eigenvalues have the largest distance in the complex 
plane~\cite{D'Ariano01,Nakahira23} as these yield the smallest overlap. We write these as $\ket{0}$ and $\ket{1}$, respectively. The action of the unitary can be 
taken to be $U\ket{0}=\ket{0}$ and $U\ket{1}=e^{i\phi} \ket{1}$ without loss of generality, where $|\phi|$ is the largest phase distance among the eigenvalues of the 
unitary. To keep track of the phase we  denote the unitary by $U(\phi)$. 

An orthonormal basis of these permutationally symmetric states is given by the well known Dicke states~\cite{Dicke:54}, 
\be
\ket{N,m}=\frac{1}{\sqrt{\binom{N}{m}}}\sum_{g\in\cS_N}\pi_g[\ket{1}^{\otimes m}\ket{0}^{\otimes(N-m)}],
\label{eq:Dicke_states}
\ee
where $\pi:\cS_N\to\mathbb{U}(2^N)$ is a permutation of the $N$ qubits, and our input state can be taken to 
be 
\be
\ket{\psi}=\sum_{m=0}^{N}\sqrt{c_m}\ket{N,m},
\label{eq:symm_state}
\ee
with $c_m\geq 0,\, \sum c_m=1$. The overlaps between the states of Eq.~\eqref{eq:unitary_discrimination} are given by 
\be
G_{kl}(\phi)=\frac{1}{N}\sum_{m=0}^Nc_m\,b_m(\phi),\; k\neq l,
\label{eq:unitary_gram}
\ee
where    
\be
b_m(\phi)=1-\frac{4m(N-m)\sin^2\nicefrac{\phi}{2}}{N(N-1)}.
\label{eq:coeff}
\ee

As $b_m(\phi)$ is independent of $k$ and $l$ the Gram matrix is again circulant and we can immediately write down the optimal 
probability of success.  The latter is maximal whenever the off-diagonal terms of the Gram matrix are minimal.  Whatever the value of 
$\phi, \, b_{\lceil\frac{N}{2}\rceil}\leq b_{m}$, with $\lceil x\rceil$ denoting the minimum integer not smaller than $x$. The latter is given 
by
\begin{equation}
b_{\lceil\frac{N}{2}\rceil}(\phi)=1-\frac{2\lceil\frac{N}{2}\rceil\sin^2\frac{\phi}{2}}{2\lceil\frac{N}{2}\rceil-1}.
\label{eq:overlap}
\end{equation}
Observe that $b_{\frac{N}{2}}(\phi)$ is a positive, monotonically decreasing function for $\phi\in[0,\phi_{\min}(N)]$, where
$
\cos(\phi_{\min}(N))=\left(-1+\frac{1}{\lceil\frac{N}{2}\rceil}\right),
$
is the value at which $b_{\lceil\frac{N}{2}\rceil}=0$.  Therefore, if $\phi\in[0,\phi_{\min}(N)]$ the optimal strategy consists of preparing 
$\ket{\psi}=\ket{N,\lceil\frac{N}{2}\rceil}$ and performing the square root measurement with the corresponding probability of success being the same as in Eq.~\eqref{eq:max_prob_succ} substituting $b$ by $b_{\lceil\frac{N}{2}\rceil}(\phi)$ given in Eq.~\eqref{eq:overlap}
\begin{widetext}
\be
P^{\mathrm{U}}_S(0\leq\phi\leq\phi_{\min}(N))=\frac{1}{N}\left(\sqrt{\cos^2\nicefrac{\phi}{2} +\xi_N\sin^2\nicefrac{\phi}{2}}+\sqrt{(N-1)(1-\xi_N)}\sin\nicefrac{\phi}{2}\right)^2,
\label{eq:perm_sym_prob_succ-1}
\ee   
\end{widetext}
where $\xi_N=0$ for $N$ even and $\xi_N=1/N^2$ for $N$ odd. Observe that the differences between the even and odd expressions are of order $O(1/N^2)$ and, 
as expected, disappear as $N$ grows large. 

Recall that Eq.~\eqref{eq:perm_sym_prob_succ-1} assumes the use of pure fully symmetric probe states. 
We have performed a numerical optimization without any assumption on the probe states using SDPs with 
see-saw techniques, where one first optimizes the measurement for a fixed state and subsequently optimizes the input state for 
the optimal measurement in the previous step. We attain the same values as those given by Eq.~\eqref{eq:perm_sym_prob_succ-1}
for up to $N=7$ systems.

Notice that something remarkable happens for $\phi_{\min}(N)<\phi\leq \pi$; for these range of values 
$b_{\lceil\frac{N}{2}\rceil}(\phi)\leq 0$ and we can exploit superpositions between symmetric 
basis states in order to identify the malfunctioning device \defn{with certainty}. 

Specifically, as $b_0(\phi)=1$ is independent of $\phi$, initializing the $N$ probes in the state 
$\ket{\psi}=\sqrt{c_0}\ket{N,0}+\sqrt{c_{\lceil\frac{N}{2}\rceil}}\ket{N,\lceil\frac{N}{2}\rceil}$ with 
\begin{equation}
c_{\lceil\frac{N}{2}\rceil}=\frac{2\lceil\frac{N}{2}\rceil-1}{2\lceil\frac{N}{2}\rceil\sin^2\frac{\phi}{2}}
\label{eq:symm_state_2}
\end{equation}
and $c_0=1-c_{\lceil\frac{N}{2}\rceil}$ guarantees that $P^{\mathrm{U}}_S(\phi_{\min}(N)<\phi\leq \pi)=1$.  

In the limit of large $N$ perfect discrimination is possible for angles $\pi-\frac{2}{\sqrt{N}}\leq \phi\leq \pi$.  
For a fixed angle $\phi$ the probability of success attains the same asymptotic expression as for source EPI, 
Eq.~\eqref{eq:max_prob_asym}, up to sub-leading order (the gap closes as at a rate equal to $\frac{\sin^2\nicefrac{\phi}{2}}{N}$).
Thus, in the limit where the number of devices is large entangling the initial probes does not enhance the probability of success.
Note, however, that for small number of devices this improvement can be quite sizable.

We would like to point out that part of our results here are reproduced in a recent publication~\cite{Hillery23}. Notwithstanding, 
this work assumes symmetric probe states of qubits {\it ab initio} and fails to notice that this probability can be made equal to one for non-trivial values of the angle 
$\phi$.
\subsection{Channel EPI} 
\label{sec:channelEPI}

We now consider the successful identification of faulty channels (Fig.~\ref{fig:EPI}(b)) described by CPTP 
maps acting on qubit systems.  Ideally, well-functioning  components implement the unitary operation $V$, with the faulty component implementing the CPTP map 
$\cF:\cB(\cH_2)\to\cB(\cH_2)$ with Kraus operator decomposition $\{F_i\}_{i=1}^r$.  As both $V$ and $\cF$ are known---only the position of the latter is 
unknown---our task reduces to discriminating among the states 
\be
\rho_k=\cI^{\otimes(k-1)}\otimes\cE\otimes\cI^{\otimes(N-k)}\otimes \cI^{\otimes N}(\proj{\psi}),
\label{eq:channel_discrimination}
\ee    
where $\cE:\cB(\cH_2)\to\cB(\cH_2)$ is a channel with Kraus operator decomposition given by 
$\{K_i=V^\dagger\, F_i\}_{i=1}^r$. $\ket{\psi}\in\cH_2^{\otimes N}\otimes\cH_2^{\otimes N}$ is the initial state of $N$ qubits and $N$ ancillas.  

For any Kraus decomposition of $\cE$ and for any input state it holds
\be
P_S(\cE,\proj{\psi})\leq \sum_{i=1}^rP_S(K_i,\proj{\psi}),
\label{eq:upper_bound}
\ee
where 
\be
P_S(K_i,\proj{\psi})=\frac{1}{N}\,\max_{\cM}\sum_{k=1}^N\tr(M_kK^{(k)}_i\,\proj{\psi} K^{(k)\dagger}_i)
\label{eq:branch_prob_succ}
\ee
denotes the optimal probability of successfully identifying the position of the action of the Kraus operator $K_i$. Equality holds in 
Eq.~\eqref{eq:upper_bound} if and only if there exists a POVM $\cM$ that optimizes all $P_S(K_i,\proj{\psi})$ simultaneously.  

Consider first the Pauli channels
\be
\cE_{\mathrm{P}}[\rho]=p_0\rho+p_1\sigma_x\rho\sigma_x+p_2\sigma_y\rho\sigma_y+p_3\sigma_z\rho\sigma_z,
\label{eq:Pauli-channels}
\ee
where  $p_1\geq p_2\geq p_3$, $(x,y,z)$ form an orthonormal Cartesian frame, $\{\sigma_x,\, \sigma_y, \, \sigma_z\}$ the standard Pauli matrices, and 
$\sum_{k=0}^3\,p_k=1$.  The rank of a Pauli channel is the number of non-zero weights $p_i$ with $i\in (1,2,3)$.
For rank-one and rank-two Pauli channels, the bound of Eq.~\eqref{eq:upper_bound} can be saturated by a simple strategy 
involving just product states and projective measurements. Indeed, for $p_3=0$ the optimal strategy corresponds to 
preparing each probe in the state $\ket{0}$ and measuring in the eigenbasis of $\sigma_z$.
Hence, for both rank-one and rank-two Pauli channels the optimal probability of success reads
\be
P_S(\cE_{\mathrm{rank-1(2)}})=1-p_0+\frac{p_0}{N}.
\label{eq:dephasing_opt}
\ee

For rank-three Pauli channels, any strategy involving $N$ probes prepared in a product state and projective measurements is 
sub-optimal.  This is because one cannot perfectly discriminate between the Pauli matrices using a single qubit.  
Indeed, the best one can hope for using such a strategy is~\cite{supple} 
\be
P^{\mathrm{sep}}_S(\cE_{\mathrm{rank}-3})=1-(p_0+p^*)+\frac{p_0+p^*}{N},
\label{eq:rank-3_lower_bound}
\ee
where $p^*=\min\{p_1,p_2,p_3\}$ and corresponds to unambiguously discriminating the most likely rank-two Pauli noise. One can, however, optimally distinguish 
among the Pauli matrices, and thus saturate Eq.~\eqref{eq:upper_bound}, by introducing $N$ ancilla qubits and preparing each 
probe-plus-ancilla in the maximally entangled state $\ket{\Phi^+}=\frac{1}{\sqrt{2}}\left(\ket{00}+\ket{11}\right)$.  
As the corresponding set of states $\{\one\otimes\sigma_i\ket{\Phi^+}\}_{i=0}^3$ are orthogonal, they can be distinguished with certainty,  and one has to make a 
random guess only in the case the identity acts.  We then recover Eq.~\eqref{eq:dephasing_opt} as the optimal probability of  success. We note that this is in 
agreement with the result of~\cite{Zhuang20a} which proved that the optimal probability of success for tele-covariant channels---of which the depolarizing channel is 
a special case---the optimal strategy for detecting the position of the channel is a non-adaptive ancilla assisted strategy.
It remains to check whether the 
upper bound to  $P_S(\cE_\mathrm{rank-3})$ is achievable without the use of ancillas.  For small network sizes, we can numerically determine the optimal probability of 
success where we observe a clear gap. By way of example, for $N=3$ and the completely depolarizing channel ($p_0=p_1=p_2=p_3=1/4$) we numerically find 
$\max_{\rho\in\cB(\cH_2^{\otimes 3})}P_S(\cE_{\mathrm{rank}-3})=0.71$, which indeed is larger than the suboptimal $2/3$ value in 
Eq.{\eqref{eq:rank-3_lower_bound}}, and smaller than the ancilla assisted value $5/6$ (Eq.~\eqref{eq:dephasing_opt}).  Just as in the case of unitary EPI we find that 
fully symmetric probe states attain the optimal probability of success.

We now consider an amplitude damping channel with Kraus operators
\be
K_0=\left(\begin{matrix} 1 &0\\0&\sqrt{1-\gamma}\end{matrix}\right)\, \quad
K_1=\left(\begin{matrix} 0 &\sqrt{\gamma}\\0&0\end{matrix}\right),
\label{eq:amplitude_damping}
\ee
with $0\leq \gamma\leq 1$ the damping parameter. The best strategy utilizing separable states can be obtained by noting that the 
action of amplitude damping on an arbitrary Bloch vector results in $\vec{r}\to(r_x\sqrt{1-\gamma},r_y\sqrt{1-\gamma},\gamma+r_z(1-
\gamma))$.  It follows that the probability of detecting the action of amplitude damping is highest if one prepares the $N$ product-state 
probes in the direction where amplitude damping is most pronounced (here in the $\hat{z}$-direction), and measuring each qubit along 
that same direction which results in 
\be
P_S^{\mathrm{sep}}(\cE_{AD})=\gamma+\frac{1-\gamma}{N}.
\label{eq:ad_lower_bound}
\ee

Clearly Eq.~\eqref{eq:ad_lower_bound} provides a lower bound to the probability of success for the most general, ancilla-assisted strategy.  An upper bound can be 
obtained from the following chain of inequalities
\begin{align} \nonumber
P_S(\cE_{AD},\rho)&\leq P_S(K_0,\rho)+P_S(K_1,\rho)\\
&\leq P_S(K_0,\rho)+p(K_1,\rho),
\label{eq:chain_inequalities} 
\end{align}
where $p(K_1,\rho)$ is the probability that the $K_1$ Kraus operator of the amplitude damping channel has acted and is 
independent of the position of the channel.  Moreover, $P_S(K_0,\rho)$ depends 
solely on the overlap of the conditional (unnormalized) pure states 
$\rho_k=\left(\one\otimes K_0^{(k)}\right)\rho\left(\one\otimes K_0^{(k)\dagger}\right)$. 

It is straightforward to check that the upper bound in Eq.~\eqref{eq:chain_inequalities} can be attained by preparing the $2N$ 
probe-ancilla systems in the state 
\be
\ket{N,m}_{pa}\equiv{{\binom{N}{m}}}^{-1/2}\sum_{g\in S_N}\Pi_g\ket{11}^{\otimes m}_{pa}\ket{00}^{\otimes N-m}_{pa},
\label{eq:double_Dicke}
\ee
where each ancilla system acts as a flag for its respective probe system and $\Pi: \cS_N\to \cH_p^{\otimes N}\otimes\cH_a^{\otimes N}$ is a representation of the 
permutation group $S_N$ acting on the total probe-plus-ancilla space , i.e., $\Pi_g=\pi_g\otimes\pi_g,\,\forall\,g\in\cS_N, \, \pi:\cS_N\to\cH_{p(a)}$.  
Observe that the total number of excitations of $\ket{N,m}_{pa}$ is even for all $m\in(0,\ldots, N)$ and that $K_0$ preserves this number, whereas $K_1$ removes 
one excitation from the probe systems but not the ancilla resulting in an odd number of total excitations.  This implies that (i) the conditional set of states 
$\left\lbrace\ket{\Psi^{(k)}_1}\right\rbrace$, that arise from the action of $K_{1}^{(k)}$ on an input state 
\be
\ket{\Psi}=\sum_{m=0}^N\sqrt{c_m}\ket{N,m}_{pa}
\label{eq:amp_damp_system_ancilla}
\ee
are orthogonal to the conditional states $\left\lbrace\ket{\Psi^{(k)}_0}\right\rbrace$ and (ii) the conditional states $\left\lbrace\ket{\Psi^{(k)}_1}\right\rbrace$ 
form an orthonormal set of states of odd number of excitations.  Properties (i) and (ii) ensure that the first and second inequalities in 
Eq.~\eqref{eq:chain_inequalities} are achievable with $p(K_1,\rho)=1-p(K_{0},\rho)={\gamma}\frac{\mean{\hat{n}}}{N}$ where $\hat{n}$ is the total excitations 
number operator.  By computing the Gram matrix for the branch corresponding to the action of $K_0$ the optimal probability of success can be 
explicitly determined to be 
\be
P_S(\cE_{AD})\leq \gamma +\frac{\sqrt{1-\gamma}(\sqrt{1-\gamma}+1)}{2 N}+O\left(N^{-2}\right),
\label{eq:asymptotic_prob_succ_ad}
\ee
and is achievable by the ancilla-assisted state with coefficients $c_{N}=p, \, c_{N-1}=1-p$ where 
\be
p=\frac{\sqrt{1-\gamma } (3 N-2)+2}{\sqrt{1-\gamma } (4 N-2)+2}.
\label{eq:opt_coeff}
\ee
We note that upper and lower bounds for the probability of success can be derived by suitably modifying the approach of~\cite{Zhuang20a} to the single shot
scenario but that these bounds are not tight and are also not demonstrably known to be achievable.

Numerically we find that for $N=3$ and $\gamma=1/2$ the optimal success probability without ancillas is  $\max_{\rho\in\cB(\cH_2^{\otimes 3})}
P_S(\cE_{AD})=0.68$---achievable by a pure fully symmetric probe state---while the ultimate limit is $P_S(\cE_{AD})=0.69$.  The difference is small, but sufficient to 
show that there is a gap in performance. Notice however, that Eq.~\eqref{eq:asymptotic_prob_succ_ad} shows that the improvement over the optimal product state 
strategy is sub-leading in $N$.  Indeed, in the limit where the number of devices to be checked is large, the gap between 
Eqs.~(\ref{eq:ad_lower_bound},~\ref{eq:asymptotic_prob_succ_ad}) closes, and it suffices to deploy our probes in 
the optimal product state. 

\section{Local Measurement Strategies for source EPI}
\label{sec:local}

In this section, we investigate four local measurement strategies and gauge their performance in determining the position of a 
malfunctioning quantum source, i.e. detecting faulty prepared states.  For ease of exposition, we will describe each of the local 
measurement strategies and state their performance only.  For a more detailed treatment of each strategy, we direct the reader to the supplementary 
material~\cite{supple}.

The simplest local strategy conceivable is to unambiguously determine the malfunctioning source.  For this, each party 
measures in the $\{\ket{0},\,\ket{1}\}$ basis.  We term such measurement strategy as the 
\defn{basic} local strategy. The probability of obtaining outcome one is $\sin^2\nicefrac{\phi}{2}$, whereas if all 
measurement outcomes are zero, which occurs with probability $\cos^2\nicefrac{\phi}{2}$, we have to guess the 
position of the malfunctioning source at random. Hence, the optimal probability of success using the basic local strategy is 
\begin{align}
P_S^{\mathrm{BL}}(\phi)=& \sin^2\left(\frac{\phi}{2}\right)+\frac{\cos^2\left(\frac{\phi}{2}\right)}{N} \nonumber \\
=& 1-\frac{(N-1)}{N} \cos^2\left(\frac{\phi}{2}\right).
\label{eq:optimal_local}
\end{align}
Notice that in the limit of $N\to\infty$ this strategy asymptotically approaches the constant value of the optimal square root 
measurement, Eq.~\eqref{eq:max_prob_succ},  but at a slower rate since the sub-leading term here is $\sim 1/N$ instead of 
$1/\sqrt{N}$ of the optimal performance. 

Next, we consider the most general local strategy involving $N$ independent measurements.  Here each party 
measures in a different basis, and for each of the $2^N$ possible measurement outcomes, $\mathbf{m}\in\{0,1\}^{N-1}$, our 
guessing rule is maximum likelihood, i.e., we choose the hypothesis, 
$k\in(1,\ldots,N)$, that maximizes the corresponding conditional probability distribution $q(\mathbf{m}|k)$,
\be
P^\mathrm{GL}_S=\frac{1}{N}\sum_{\mathbf{m}=0}^{2^N-1} \max_k\{q(\mathbf{m}|k)\}.
\label{eq:general_local}
\ee
The best such strategy corresponds to maximizing Eq.~\eqref{eq:general_local} over the $N$ local 
measurement basis. 

Our third local strategy is a \defn{greedy} strategy, a sequential strategy that uses forward 
communication in order to optimally distinguish among the most likely hypothesis at every step.
In this strategy each party $n\in\{1,\ldots, N\}$, starting with $n=1$, performs a measurement 
on the $n$-th subsystem of $\ket{\psi_k}\in\cH^{\otimes N}$ corresponding to their location and   
attempts to identify the location of the error. As we show in~\cite{supple} the optimal
strategy for each party $n$ corresponds to a binary hypothesis testing scenario 
where the hypotheses are that the error is at location $n$, or at the most likely position among
the remaining hypotheses given the prior information at party $n$'s disposal,  by means of the prior distributions $p^{(n)}(k), \ k\in(1, \ldots N)$. These probabilities 
are updated sequentially, using Bayes' rule, depending on the outcomes of all parties up to and including
party $n-1$.  Party $N$ then measures their subsystem using the optimal measurement derived from the prior distribution $p^{(N)}(k)$.  The corresponding probability 
of success, which depends on the measurement record $\mathbf{m}\in\{0,1\}^{N-1}$, is $P^{(N)}_S(\mathbf{m})$ (see
Eq.~\eqref{eq:Greedy_success}) and the average probability of success is given by 
\be
P^\mathrm{Gr}_S=\sum_{\mathbf{m}}q(\mathbf{m})\, P^{(N)}_S(\mathbf{m}) ,
\label{eq:greedy_local}
\ee 
where $q(\mathbf{m})$ denotes the probability of obtaining the measurement record $\mathbf{m}\in\{0,1\}^{N-1}$. 

Our last local measurement strategy is again a sequential adaptive strategy where we fix our guess for the most likely hypothesis 
to be the position of the last positive outcome in any given measurement record (the \defn{last one} strategy).  By way of example 
suppose that $N=3$.  Then for the measurement records $\{001,\, 011,\, 101,\, 111\}$ we hypothesize that the error occurred at 
position $k=3$, for the measurement records $\{010,\, 110\}$ that the error occurred at $k=2$, for $\{100,\, 000\}$ that the error 
occurred at $k=1$.  Notice that here we need to optimize over $2^N-1$ parameters, since as the optimal measurement basis at 
each position $n$ depends on all previous $k-1$ measurement outcomes.  The corresponding probability of success is given 
by 
\be
P_S^\mathrm{L1}=\frac{1}{N}\sum_{k=1}^N \,\sum_{\bm m\in S_k}\, q(\mathbf{m}|k),
\label{eq:probsucc_ramon}
\ee 
where 
\be
q(\mathbf{m}|k)=\prod_{i=1}^N\lvert\braket{m_i}{\psi_k}\rvert^2,
\label{eq:conditionals_ramon}
\ee
and $S_k\equiv\{\mathbf{m}\lvert m_{k+1}\ldots m_N=0\}$.

\begin{figure}[htb]
\includegraphics[width=8cm]{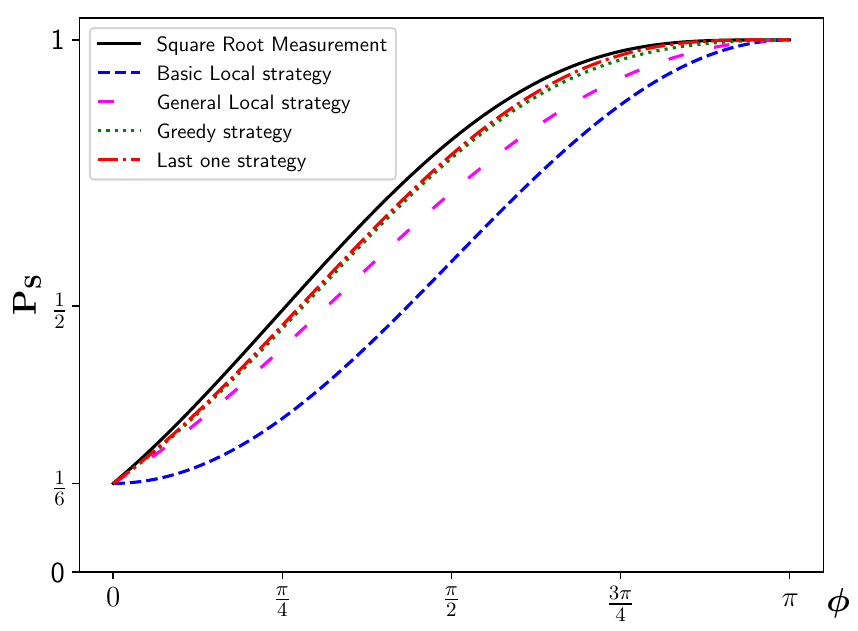}
\caption{The probability of success for EPI of sources using the square root measurement (solid black line), the basic local strategy (blue dashed line), the greedy 
strategy (green dotted line), the general local strategy (magenta dashed line), and the last one strategy (red dot-dashed line).  All strategies are evaluated for the case 
of $N=6$.}
\label{fig:localcomparisons}
\end{figure}

The performance of all four strategies for the case where $N=6$ is shown in Fig.~\ref{fig:localcomparisons}. The worst 
performing strategy is the basic local strategy, with the general local,  greedy, and last one strategies performing much better.  
This is hardly surprising since adaptive strategies utilizing information from previous measurement outcomes are expected to 
perform better.  Moreover, the last one strategy performs best as it utilizes the entire past measurement record, contrary to the 
greedy strategy which only utilizes the information from the previous measurement.  While the improvement of the last one 
strategy over the greedy one is only slight---the corresponding probabilities of success differ only in the third digit---we have 
checked that this improvement does persist also for moderate network sizes, and that there is a clear non-zero gap with the optimal 
measurement strategy employing the square root measurement. 

Notice that, to zeroth order in $N$, all four strategies achieve the maximum probability of success in the large $N$ limit, since as 
$\lim_{N\to\infty}P^\mathrm{BL}_S(\phi)=\lim_{N\to\infty}P^*_S(\phi)=\sin^2\nicefrac{\phi}{2}$ and 
$P^\mathrm{BL}_S<P^\mathrm{GL}_S<P^\mathrm{Gr}_S<P^\mathrm{L1}_S<P^*_S$. The differences between all strategies is the speed with which one approaches 
this constant.  As shown in  Eq.~\eqref{eq:optimal_local} the basic local strategy approaches the asymptotic value from above with a scaling of $N^{-1}$, i.e.  the 
success probability decreases at a faster rate than the optimal strategy, which shows a  $N^{-1/2}$ scaling [Eq.~\eqref{eq:max_prob_asym}]. We can only assess the 
performance of the other local strategies by numerical optimization, which becomes exceedingly hard for large values of $N$. The most tractable case is that of the 
greedy strategy for which we have computed $P_S^\mathrm{Gr}(N)$ up to $N=20$, exhibiting a scaling that is compatible with $N^{-2/3}$.

It is worth stressing that, aside from being easily implementable, there are other advantages that one 
needs to consider when choosing among local strategies.  For example, the basic local 
strategy may identify the position of an error with high confidence in an online fashion, even in settings where the number of samples $N$, is not fixed in advance.
The use of quantum sequential methodologies in settings without a finite horizon has been recently studied 
in~\cite{vargas_quantum_2021,li_optimal_2022,fanizza_ultimate_2023,gasbarri_sequential_2023}.

\section{Conclusions}  

We have addressed how to optimally identify the position of a malfunctioning quantum device that forms 
part of an interconnected quantum network in the simplified case where the latter consists of $N$ identical devices that can be 
addressed in a parallel fashion.  For unitary EPI we discover that entanglement enhances the probability of correct 
identification, and even allows for perfect identification of the device if the unitary rotation angle is greater than 
some threshold.  For rank-one and rank-two Pauli channels the optimal strategy involves separable states and measurements, 
whereas for rank-three and amplitude damping channels we find that the optimal identification strategy requires 
entanglement with $N$ additional ancillas.  However, the use of entanglement among the input probes only pays dividends if the number 
of devices needed to be checked is small; as the network size grows large strategies employing separable states do just as well.

We have numerically verified that for unitary EPI, ancilla-free totally symmetric pure states attain the optimal performance. While we have given a formal proof that 
ancillas are not necessary to achieve optimality, the sufficiency of pure permutationally symmetric states remains unproven.
In channel identification using probes without ancillas, we find again that symmetric pure states stand as the optimal choice.  Although the numerical evidence in both 
cases is very compelling, an analytical proof presents an intriguing open challenge of theoretical interest.

There are several interesting extensions of the current work that are of both theoretical as well as practical significance. 
For instance, in this work we have assumed throughout that devices act on \emph{separate} systems.  In a quantum computer, or a 
general quantum network the same quantum system may be subject to a \emph{sequence} of quantum operations. It is natural then to 
ask how to optimally identify malfunctioning devices that act in sequence on the same quantum system.  Furthermore, it is interesting to 
know what is the optimal performance in cases where both the position as well as the identity of the anomaly are required. 

\section{Acknowledgements}
This work was supported by MCIN/AEI/10.13039/501100011033  with projects PID2022-141283NB-I00 with the support of FEDER funds and PID2021-128970OA-
I00, with funding from European Union NextGenerationEU (PRTR-C17.I1) and  Generalitat de Catalunya. We also acknowledge financial support from the Ministry for 
Digital Transformation and of Civil Service of the Spanish Government through the QUANTUM ENIA project call Quantum Spain project, and by the European Union 
through the Recovery, Transformation and Resilience Plan NextGeneration EU within the framework of the Digital Spain 2026 agenda. MS also acknowledges funding 
from the Ministerio de Ciencia e Innovación of the Spanish Government under the Ramon y Cajal funding scheme (RYC2021-032032-I) and FEDER funds 
C.EXP.256.UGR23 from the Junta Andalucia. JC also acknowledges support from the ICREA Academia award, and form the QuantERA project C’MON-QSENS!, by 
Spanish MICINN PCI2019-111869-2.

\bibliographystyle{unsrt}
\bibliography{pei}
\clearpage
\setcounter{equation}{0}
\appendix
\section{Optimal states for unitary EPI require no ancillas}
\label{app:no-ancillas}
In this appendix, we provide the proof that for the case of unitary EPI, the optimal probability of success can be achieved without ancillas.  To that end let us write the malfunctioning unitary of dimension $d$ as 
\be
U=\sum_{a=0}^{d-1} e^{\ii\phi_a}\proj{a}
\label{app:mafunctioningU}
\ee
where $\{\ket{a}\}_{a=0}^{d-1} $ are the $d$ eigenvectors of the unitary $U$.  Using the same orthonormal basis we may write any probe-plus-ancilla state as 
\be
\ket{\psi}=\sum_{\bf m,n=0}^{d^N-1}c_{\bf nm}\ket{\bf n}\otimes\ket{\bf m}
\label{app:probe-plus-ancilla}
\ee  
where ${\bf n},\,{\bf m}$ denote $N$ $d$-nary strings.  The elements of the Gram matrix are then given by 
\begin{align}\nonumber
G_{kl}&=\sum_{\bf n,n',m=0}^{d^N-1}c^*_{\bf nm}c_{\bf n'm}\bra{\bf n}U^\dagger_k U_l\ket{\bf n'}\\  \nonumber
&=\sum_{\bf n,m=0}^{d^N-1}\left|c_{\bf nm}\right|^2e^{-\ii(\phi_{n_k}-\phi_{n_l})}\\
&=\sum_{\bf n=0}^{d^N-1}\left|\psi_{\bf n}\right|^2e^{-\ii(\phi_{n_k}-\phi_{n_l})}
\label{app:Gram-matrix}
\end{align}
where $\left|\psi_{\bf n}\right|^2=\sum_{\bf m=0}^{d^N-1}\left|c_{\bf nm}\right|^2$.  But this is precisely the same Gram matrix element one would obtain using the no-ancilla state
\be
\ket{\psi}=\sum_{\bf n=0}^{d^N-1}\psi_{\bf n}\ket{\bf n},
\label{app:N-qubit state}
\ee
proving the claim.

\section{Optimality of permutational invariant states and covariant measurements}
\label{app:perm_inv}
In this appendix we prove that given any state $\rho=\ketbra{\psi}{\psi}$ and POVM $\{M_k\}$ there exists a permutationally invariant 
state $\tau$ and covariant POVM that achieves the same probability of success. 

Recall that the probability of successfully identifying the position of a malfunctioning device is given by 
\be
P_S(\cE,\proj{\psi})=\max_{\{E_k\geq0\}}\,\frac{1}{N}\sum_{k=1}^N\tr\left(\cE_k\left(\proj{\psi}
\right)E_k\right),
\label{app:prob_succ}
\ee
where $\ket{\psi}\in\cH_2^{\otimes N}$ and $\cE_k:\cB(\cH_2^{(k)})\to\cB(\cH_2^{(k)})$.  Alternatively, we may think of the position 
of the channel as fixed, acting only on the last qubit, with a pre- and post-processing of the input and output states by the shift 
superoperator $\cT^{N-k}(\cdot)=T^{N-k} (\cdot)\, T^{-(N-k)}$, where $T:\Z_N\to\mathbb{U}(2^N), \, T^N=\one$ a unitary 
representation of $\Z_N$, i.e, 
\be
\cE_k=\cT^{-(N-k)}\circ\cE_N\circ\cT^{N-k},
\label{app:trans_symmetry}
\ee
with $\cT\circ\cE(A)=\cT(\cE(A))$.  Moreover, as we are promised that only a single error occurs, we can extend the translation 
symmetry of the channel to the full permutation group $S_N$ by noting that 
\be
\cT^{N-k}(\cdot)=\frac{1}{N!}\sum_{\sigma\in S_N|N\stackrel{\sigma}{\to}k}\pi_\sigma(\cdot)\pi^\dagger_\sigma,
\label{app:trans_to_perm}
\ee
where $\pi:S_N\to \mathbb{U}(2^N)$ is a unitary representation of $S_N$.  

Eq.~\eqref{app:trans_to_perm} simply states that the translation of the channel from position $N$ to $k$ is equivalent to the 
$(N-1)!$ permutations that map position $N$ to $k$.  We can also modify our search over the optimal POVM accordingly by
associating all $(N-1)!$ measurement outcomes $E_\sigma$ that map $N$ to $k$ so that Eq.~\eqref{app:prob_succ} reads
\begin{align}
P_S(\cE,\proj{\psi})=& \max_{\{E_\sigma\geq0\}}\, \frac{1}{N} \nonumber \\
&\times \sum_{\sigma\in S_N}
\tr\left(\cV^\dagger_\sigma\circ\cE_N\circ\cV_\sigma\left(\proj{\psi}\right)\,E_\sigma\right),
\label{app:prob_succ_perm}
\end{align}
where $\cV_\sigma(\cdot)=\pi_\sigma(\cdot)\pi^\dagger_\sigma$ and $\cV^\dagger_\sigma=\cV^{-1}_{\sigma}=\cV_{\sigma^{-1}}$.  

For a fixed, equiprobable set of quantum states $\{\rho_k\}_{k=1}^N$ the optimal probability of successful discrimination 
can be written as the following semi-definite program (SDP) in dual form~\cite{Holevo73,Yuen75,Helstrom76,Watrous} 
\begin{align}\nonumber
&\min_{\Gamma\geq0}  \tr\,\Gamma\\ 
\mathrm{subject\,to}\quad&\Gamma\geq \frac{1}{N}\rho_k\quad \forall\, k\in(1,\ldots,N)\,.
\label{eq:Holevo_conditions}
\end{align}
Using Eq.~\eqref{eq:Holevo_conditions} and the properties of the trace we have 
\begin{align}\nonumber
\tr\, \Gamma&=\tr\left(\pi_\sigma\,\Gamma\,\pi^\dagger_\sigma\right)\quad \forall\, \sigma\in S_N\\    \nonumber
&=\frac{1}{N!}\tr\left(\sum_{\sigma\in S_N}\,\pi_\sigma \Gamma\, \pi^\dagger_\sigma\right)\\
&\equiv \tr\,\tilde{\Gamma},
\label{app:perm_symm_Lagrange}
\end{align}
where $\tilde{\Gamma}$ is a permutationally invariant operator. Furthermore, the optimality conditions of 
Eq.~\eqref{eq:Holevo_conditions} imply
\begin{align}\nonumber
&\Gamma\geq \frac{1}{N!}\,\pi^\dagger_\sigma\, \cE_N(\pi_\sigma\proj{\psi}\pi^\dagger_\sigma)\pi_\sigma, \;\;\forall\,\sigma\in 
S_N\\  \nonumber
&\pi_\sigma\,\Gamma\, \pi^\dagger_\sigma\geq  \frac{1}{N!}\, \cE_N(\pi_\sigma\proj{\psi}\pi^\dagger_\sigma),\;\;\forall\,\sigma\in 
S_N\\
&\tilde{\Gamma}\geq \frac{1}{N!}\,\cE_N(\cG[\proj{\psi}]),
\label{app:perm_symm_Holevo}
\end{align}
where $\cG[\proj{\psi}]=\frac{1}{N!}\sum_{\sigma\in S_N}\,\pi_\sigma\proj{\psi}\,\pi^\dagger_\sigma$, and has the property of 
symmetrizing $\proj{\psi}$. 

As it is sufficient to restrict to permutationally invariant states Eq.~\eqref{app:prob_succ_perm} reduces to 
\begin{align}\nonumber
P_S(\cE,\tau)&=\max_{\{E_\sigma\geq0\}}\,\max_{\{\tau\}}\frac{1}{N}\sum_{\sigma\in S_N}\tr\left(\cE_N(\tau)\cV^\dagger_\sigma(E_\sigma)\right)\\
&=(N-1)!\max_{\{E\geq0\}}\,\max_{\{\tau\}}\tr\left(\cE_N\left(\tau)\right)E\right),\,
\label{app:prob_succ_perm_state}
\end{align}
where, the second maximization is over all $\tau$ that satisfy $[\pi_\sigma,\tau]=0,\forall\,\sigma\in S_N$, and we have made use of the fact that $\{E_\sigma=\pi_\sigma E \pi^\dagger_\sigma\}$ constitutes a covariant measurement 
whose fiducial element is $E$, and used the identity $\cG[E]=\frac{1}{N!}\sum_{\sigma\in S_N}\cV_\sigma(E)=\frac{1}{N!}\one$.

Eq.~\eqref{app:prob_succ_perm_state} tells us that for any given input state $\ket{\psi}$ and optimal 
POVM $\{M_k\}$ there exists a permutationally symmetric state $\tau$ and permutationally covariant POVM 
$\{E_\sigma=\pi_\sigma E \pi^\dagger_\sigma\}$ that achieves the same probability of success.  Moreover, the optimization 
over the covariant measurement requires us to optimize over a single \emph{fiducial} POVM element $E$ such that 
\be
N!\,\cG(E)=\one.
\label{app:cov_meas_constraint}
\ee

\section{Optimal product state strategy for rank-3 Pauli channels}
\label{app:rank3noise}

In this appendix we show that the optimal, ancilla-free, product state strategy for detecting rank-3 Pauli noise using projective 
measurements has a probability of success given by Eq.~\eqref{eq:rank-3_lower_bound}.  To do so we will first show that the 
optimal probability of success can be achieved by unambiguously determining the position of the error channel and then 
show that the optimal unambiguous strategy is indeed given by Eq.~\eqref{eq:rank-3_lower_bound}

Consider then the case where each party measures its corresponding system locally using a fixed, non-adaptive but otherwise 
completely arbitrary measurement.  Without loss of generality let use denote the measurement outcomes of each party's 
measurement by 0 and 1.  In a completely arbitrary fashion let us also make the assignment that outcome zero means that 
the channel did not act whereas outcome one means the channel has acted.  After all parties have performed their 
measurement we will obtain one of the $2^N$ possible measurement outcomes $\{{\bf m}\equiv m_1\ldots,m_N\,\vert\, 
m_i\in(0,1),\, \forall\, i\in(1,\ldots,N)\}$.

For the outcome ${\bf 0}$ we simply guess at random the position of the channel, whereas for the $N$ 
measurement outcomes with Hamming weight one, our guess for the position of the channel is that it acted at the position for 
which the measurement outcome $m_i=1$.  For example, if $m_1=1,\, m_i=0,\, \forall\, i\in(2,\ldots,N)$ we guess that the 
channel acted on the first qubit.  Our probability of success is then given by 
\be
\begin{split}
P_S&=p(g=1|k=1)p(k=1)\Pi_{i=2}^Nq(i=0|i=0)\\
&\equiv\frac{1}{N}p(g=1|k=1)q^{N-1}.
\end{split}
\label{app:Hamming1position1}
\ee
Here $p(g=i|k=i)=\tr(\ket{1}_i\bra{1}\cE(\ket{\psi}_i\bra{\psi}))$ denotes the probability that $m_i=1$ given that the
channel, $\cE$, acted at position $i$ (i.e., correctly identifying the channel), $1-p(g=i|k=i)$ is the conditional probability that 
$m_i=0$ given the channel acted at position $i$ and $q(i=0|i=0)$ is the conditional probability that $m_i=0$ given the state at 
position $i$ is $\ket{\psi}$ (i.e., correctly identifying that the channel did not happen at position $i$) with $1-q(i=0|i=0)$ denoting 
the conditional probability that $m_i=1$ given the state at position $i$ is $\ket{\psi}$. As all the latter probabilities are the same we simply drop the label $i$ in what follows and write $q$ for simplicity. It follows that if the measurement record 
contains a single element different from zero at position $i$ the probability of successfully identifying the channel is given by 
\be
P_S=\frac{1}{N}p(g=i|k=i)q^{N-1}.
\label{app:Hamming1}
\ee 

Now consider a measurement record with Hamming weight $1\leq r\leq N$, and suppose that for such measurement outcome 
we randomly choose one of the $r$ positions $\{i\in(1,\ldots,r)\,\vert m_i=1\}$ as our guess.  The probability of 
success now reads
\begin{align}\nonumber
P_S&=\frac{1}{N}\sum_{i=1}^Np(g=1|k=1)\sum_{r=1}^N\binom{N-1}{r-1} \frac{q^{N-r}(1-q)^{r-1}}{r}\\   \nonumber
&=p(g=1|k=1)\sum_{r=1}^N\binom{N-1}{r-1} \frac{q^{N-r}(1-q)^{r-1}}{r}\\   \nonumber
&=\frac{p(g=1|k=1)}{N}\sum_{r=1}^N\binom{N}{r} q^{N-r}(1-q)^{r-1}\\   \nonumber
&=\frac{p(g=1|k=1)}{N}\frac{1}{1-q}\sum_{r=1}^N\binom{N}{r} q^{N-r}(1-q)^{r}\\   \nonumber
&=\frac{p(g=1|k=1)}{N}\frac{1}{1-q}\left(\sum_{r=0}^N\binom{N}{r} q^{N-r}(1-q)^{r}-q^N\right)\\
&=\frac{p(g=1|k=1)}{N}\frac{1-q^N}{1-q},
\label{app:Psgeneral}
\end{align}
where we have used the fact that $p(g=i|k=i)$ is the same for all $i\in(1,\ldots,N)$ in going from the first to the second line in 
Eq.~\eqref{app:Psgeneral}. For $q<1$ the success probability scales as $\frac{1}{N}$ whereas for $q=1$, i.e., for unambiguously discriminating when the channel acted, the success probability is given by 
\be
P_S=p(g=1|k=1)
\label{app:unambiguous}
\ee
It follows that the optimal minimum error strategy for product states with projective measurements is the one that 
unambiguously detects the action of the channel.  

All that is left now is for us to determine the initial state $\ket{\psi}$ and corresponding unambiguous measurement strategy 
that maximizes $p(g=1|k=1)$.  Looking at the definition of rank-3 Pauli channels (Eq.~\eqref{eq:Pauli-channels}) it follows that 
the maximum of $p(g=1|k=1)$ occurs by tailoring the initial state $\ket{\psi}$ such that we can distinguish with certainty the 
action of two out of the three Pauli operators with the largest probability of occurrence, and randomly guessing the position of 
the channel for the remaining Pauli operator and the identity.  Denoting by $p^*=\min\{p_1,p_2,p_3\}$ it follows that 
\be
P_S=1-(p_0+p^*)+\frac{p_0+p^*}{N}.
\label{app:final_result}
\ee

Interestingly enough one may arrive at the conclusion that the optimal probability of success is achievable by unambiguous 
discrimination by employing the following strategy.  For any measurement record 
other than $\bf 0$---for which we simply guess at random---we nominate as our guess for the position of the channel to be that 
corresponding to the last one in the measurement record.  For example, for $N=3$ we nominate the position of the channel to 
be $k=2$ for both measurement records $010$ and $110$.  The probability that we successfully identify the position of the 
channel is given by 
\be
\begin{split}
P_S&=p(g=2|k=2)q^2+p(g=2|k=2)q(1-q)\\
&=p(g=2|k=2)q\, ,
\end{split}
\ee
which is clearly maximal if $q=1$.  More generally this strategy yields for the optimal probability of success
\be
P_S=\frac{1}{N}\sum_{i=1}^Np(g=i|k=i)q^{N-i}
\label{app:last1prob}
\ee
which is clearly optimized by an unambiguous strategy, namely $q=1$.  Notice the remarkable simplicity of the argument by cleverly choosing our guessing strategy for each measurement record.
\section{Local Measurement Strategies}
\label{app:local}

In this appendix we provide the details for all local strategies described in Sec.~\ref{sec:local}.

\emph{General (fixed) local strategy:}
We begin with the most general strategy involving $N$ independent fixed  measurements. Here qubit $i$ is measured in the basis  
\be
\ket{m_i}=\cos\left(\frac{\theta_i+m_i\pi}{2}\right)\ket{0}+\sin\left(\frac{\theta_i+m_i\pi}{2}\right)\ket{1},
\label{app:measurement_basis}
\ee
where $m_i\in(0,1)$.  The conditional probability of obtaining any of the $2^N$ measurement records $\{\mathbf{m}\}$, given the 
state is $\ket{\psi_k}=\ket{0}^{\otimes k-1}\otimes\ket{\phi}\otimes\ket{0}^{\otimes N-k}$ is   
\begin{eqnarray}
q(\mathbf{m}|k)&=\prod_{i=1}^{k-1}\cos^2\left(\frac{\theta_i+m_i\pi}{2}\right)\cos^2\left(\frac{\theta_k-\phi+m_k\pi}{2}\right)\nonumber \\
&\times \prod_{i=k+1}^N\cos^2\left(\frac{\theta_i+m_i\pi}{2}\right).
\label{app:cond_prob_bl}
\end{eqnarray}
Upon obtaining a given measurement record we need to assign a guess as to which one of the $N$ hypothesis 
$k\in(1,\ldots,N)$ is the most likely one. This is achieved by choosing the hypothesis that maximizes the corresponding 
posterior probability distribution, $\{p(k)q(\mathbf{m}|k)\}_{k=1}^N$.  Hence, the resulting probability of success is given by 
\be
P^\mathrm{GL}_S=\sum_{\mathbf{m}=0}^{2^N-1} \max_k\{p(k) q(\mathbf{m}|k)\}.
\label{app:max_prob_bl}
\ee
The maximum probability of success under the general local strategy corresponds to optimizing over the $N$ independent measurement angles $\{\theta_i\}$ which can be done efficiently using numerical techniques.

\emph{Greedy strategy:}
Next we consider a greedy local strategy that uses forward communication from each measurement to the 
next.  Specifically, we imagine a total of $N$ parties each of which has access to a corresponding subsystem of the state 
$\ket{\psi_k}\in\cH^{\otimes N}, \, k\in\{1, \ldots, N\}$. The strategy proceeds sequentially with the first party performing a measurement
on their part of the state $\ket{\psi_k}$ communicating the outcome to the second party and so on. Each party aims to maximize the probability of 
successfully identifying the position of the error based on the information received from the previous party and the information obtained from their 
measurement. The prior information available to party $n$ is encapsulated in the prior probabilities for each hypothesis 
$\{p^{(n)}(k\vert \mathbf{m})\}_{k=1}^N$, where to ease the notation,
$\mathbf{m}$ is understood to run over the first $n-1$ outcomes, i.e., the measurement record $\{m_1,\ldots,m_{n-1}\}$. 

Party $n$ now attempts to locate the position of the error by assigning a POVM element $E_m$ to 
each of the $m=1,\ldots, N$ possible hypotheses. The corresponding probability of success is given by 
 \begin{widetext}
   \begin{align}
P^{(n)}_S (\mathbf{m}) & =\tr[E_n\rho_\phi] p^{(n)}(n|\mathbf{m})+\sum_{k\neq n} \tr[E_k\rho_0] p^{(n)}(k|\mathbf{m})
\leq \tr[E_n\rho_\phi] p^{(n)}(n|\mathbf{m})+ p^{(n)}(k^*|\mathbf{m})\sum_{k\neq n} \tr[E_k\rho_0]\nonumber \\
=& \tr[E_n\rho_\phi] p^{(n)}(n|\mathbf{m})+ p^{(n)}(k^*|\mathbf{m})\tr[(\one-E_n)\rho_0]
=p^{(n)}(k^*|\mathbf{m})+\tr[E_n (p^{(n)}(n|\mathbf{m})\rho_\phi -p^{(n)}(k^*|\mathbf{m})\rho_0)]\nonumber \\
\leq & \frac{p^{(n)}(n|\mathbf{m})+p^{(n)}(k^*|\mathbf{m})}{2} 
+ \tfrac{1}{2} \|p^{(n)}(n|\mathbf{m}) \rho_\phi -p^{(n)}(k^*|\mathbf{m})\rho_0\|_1 \, ,
\label{eq:Greedy_success}
\end{align} 
\end{widetext}
where $k^*=\mathrm{argmax}_k \{p^{(n)}(k|\mathbf{m})\}_{k\neq n}$, with the first inequality attained by picking 
$E_m=0$ for $m\neq\{n,k^*\}$, and $E_{k^*}=\one-E_n:=E_0$, and the second inequality attained by choosing $E_n$ to be the projection onto 
the positive eigenspace of $p^{(n)}(n|\mathbf{m}) \rho_\phi -p^{(n)}(k^*|\mathbf{m})\rho_0$, where $\rho_0=\ketbra{0}{0}$ and $\rho_\phi=\ketbra{\phi}{\phi}$, respectively.

That is, party $n$
will always guess in favor of either the error being at location $n$ (corresponding to the measurement outcome that we henceforth label by $m_n=1$) or at the most likely alternative location $k^*$ (corresponding to measurement now labeled $m_n=0$). 
As this is true 
for all parties $n\in\{1, \ldots, N\}$ the corresponding measurement record after party $n$ measures is 
one out of $2^n$ possible binary strings, i.e., $\mathbf{m}\in\{0, 1\}^{n}$.  

The probability that party $n$ obtains the measurement outcome $m\in\{0, 1\}$ is given by 
\begin{equation}
q(m_n)=\tr\left(E_{m_n}\left(p^{(n)}(k|\mathbf{m})\rho_\phi+(1-p^{(n)}(k|\mathbf{m}))\rho_0\right)\right)\, .
\end{equation}
Upon obtaining outcome $m_n$, we use Bayes' rule to update the priors to,     
\begin{equation}
p^{(n+1)}(k\vert\mathbf{m})=\begin{cases}
  \frac{\tr(E_{m_n} \rho_\phi)p^{(n)}(k|\mathbf{m})}{q(m_n)} & \text{for}\, k=n\\
            \frac{\tr(E_{m_n} \rho_0)p^{(n)}(k|\mathbf{m})}{q(m_n)} & \text{otherwise}
        \end{cases}\, 
\end{equation}
which are then used by party $n+1$ to pick the optimal measurement accordingly. This process is iterated 
until the last party $N$ is reached. The success probability after all $N$ samples have been measured is 
given by $P_S^{(N)}(\mathbf{m})$. Hence, the average probability of success is 
\be
P^\mathrm{Gr}_S=\sum_{\mathbf{m}=0}^{2^{N-1}-1} q(\mathbf{m}) P_S^{(N)}(\mathbf{m})
\label{eq:PSGreed}
\ee
where 
\be
q(\mathbf{m}) = \prod_{n=1}^{N-1} q(m_n)\, .
\ee
It is important to note that the greedy strategy described above is capable of  
providing a guess as to the location of the error even if, for some reason, only the first $n<N$ parties 
are able to perform measurements.

\emph{Last-one strategy:}
Finally, let us analyze a local measurement strategy that uses a guessing rule other than maximum likelihood. Specifically, we choose as our guess the hypothesis corresponding to the last one in the measurement record, and 
optimize over the measurement angles for each of the $N$ measurements.  For $\mathbf{m}=\bm 0$ we 
nominate the first hypothesis as our guess.

The first von Neumann measurement is parametrized by a single angle $\theta_1(0)$ such that 
\be
\ket{m_1=a}=\cos\left(\frac{\theta_1(a)}{2}\right)\ket{0}+\sin\left(\frac{\theta_1(a)}{2}\right)\ket{1},
\ee
where $\theta_1(1)=\theta_1(0)+\pi$, are the corresponding projection operators.  Depending on the measurement outcome 
$m_1$ the second measurement is parametrized by two angles $\theta_{2}(0|m_1),\,m_1\in(0,1);\,\theta_{2}(1|m_1)=\theta_{2}(0|
m_1)+\pi$, the third by four angles and so on. For $N$ hypothesis the total number of measurement angles that we need to 
optimize are $2^N-1$, and  corresponding probability of success is given by 
\be
P_S^\mathrm{L1}=\sum_{n=1}^Np(n) \,\sum_{\mathbf{m}\in S_n}\, q(\mathbf{m}|n),
\label{app:probsucc_ramon}
\ee 
where 
\be
q(\mathbf{m}|n)=\prod_{i=1}^N\lvert\braket{m_i}{\psi_n}\rvert^2,
\label{eq:conditionals_ramon-2}
\ee
and $S_n\equiv\{\mathbf{m}\lvert m_{n+1}\ldots m_N=0\}$.  
\end{document}